\documentclass[aps,pra,twocolumn,floatfix,showpacs,superscriptaddress]{revtex4-1}

\usepackage{color}
\usepackage{graphicx}

\begin{document}
\title{Cold atomic and molecular collisions: approaching the universal loss regime}

\author{Matthew D. Frye}
\affiliation{Joint Quantum Centre (JQC) Durham-Newcastle, Department of
Chemistry, Durham University, South Road, Durham DH1 3LE, United Kingdom}
\author{Paul S. Julienne}
\affiliation{Joint Quantum Institute, University of Maryland and NIST, College
Park, Maryland 20742, USA}
\author{Jeremy M. Hutson}
\affiliation{Joint Quantum Centre (JQC) Durham/Newcastle, Department of
Chemistry, Durham University, South Road, Durham DH1 3LE, United Kingdom}

\begin{abstract}
We investigate the behaviour of single-channel theoretical models of cold and
ultracold collisions that take account of inelastic and reactive processes
using a single parameter to represent short-range loss. We present plots of the
resulting energy-dependence of elastic and inelastic or reactive cross sections
over the full parameter space of loss parameters and short-range phase shifts.
We then test the single-channel model by comparing it with the results of
coupled-channel calculations of rotationally inelastic collisions between LiH
molecules and Li atoms. We find that the range of cross sections predicted by
the single-channel model becomes increasingly accurate as the initial LiH
rotational quantum number increases, with a corresponding increase in the
number of open loss channels. The results suggest that coupled-channel
calculations at very low energy (in the s-wave regime) could in some cases be
used to estimate a loss parameter and then to predict the range of possible
loss rates at higher energy, without the need for explicit coupled-channel
calculations for higher partial waves.
\end{abstract}

\date{\today}

\maketitle

\section{Introduction}

Elastic and inelastic or reactive collisions play a crucial role in cold atomic
and molecular gases \cite{Weiner:RMP:1999,Chin:RMP:2010}. Elastic collisions
are used for cooling and coherent control of cold gases, whereas inelastic or
reactive collisions remove atoms or molecules from traps and limit the lifetime
of cold samples. Inelastic or reactive collisions have been observed in thermal
samples at temperatures as low as 200~nK \cite{Ospelkaus:react:2010}, and cold
collisions can also be studied using merged \cite{Henson:2012} or
decelerated~\cite{vandeMeerakker:2012} beams at collision energies $E$
corresponding to tens of mK to 1~K.

In relatively simple systems, such as pairs of alkali-metal atoms or light
atom-molecule systems, it is feasible to solve the many-dimensional
Schr\"odinger equation directly using coupled-channel methods. However, for
heavier and more complex systems, the number of channels required for
convergence is too large and coupled-channel methods become prohibitively
expensive. In this regime, considerable success has been achieved with
effective single-channel methods that take account of short-range loss, whether
inelastic or reactive, with a single parameter. In particular, Idziaszek and
Julienne \cite{Idziaszek:PRL:2010}, Kotochigova \cite{Kotochigova:react:2010}
and Gao \cite{Gao:react:2010} have developed approaches based on quantum defect
theory (QDT), which takes advantage of the fact that the short-range
wavefunction is only weakly dependent on energy near threshold
\cite{Mies:MQDT:1983, Julienne:1989, Gao:2008}. If the interaction potential
$V(r)$ has an inverse power dependence on the interspecies distance $r$ at long
range, $V_{\rm LR}(r)=-C_n/r^n$, the long-range wavefunction may be expressed
in terms of the analytical solutions for the long-range potential
\cite{Gao:2008,Gao:ion:2010}. The model of Idziaszek and Julienne
\cite{Idziaszek:PRL:2010} successfully explained the temperature dependence of
reactive KRb+KRb collisions at temperatures below 1~$\mu$K, and was later
extended \cite{Idziaszek:PRA:2010} to handle the additional $r^{-3}$
dipole-dipole potential that exists when the KRb molecules are oriented with an
external electric field. More recently, Jachymski {\em et al.}
\cite{Jachymski:react:2013, Jachymski:react:2014} have extended similar models
up to the high-temperature limit, and used them to interpret merged-beam
experiments on Penning ionisation in collisions of metastable He with Ar
\cite{Henson:2012}.

The single-channel models can be expressed in terms of two parameters. One of
these describes the probability that loss will occur when the particles reach
short range, while the second describes a short-range phase shift that
characterises a background scattering length for the interaction. In the limit
of complete loss at short range, the loss rate is independent of the background
scattering length; this has been termed the ``universal" limit
\cite{Idziaszek:PRL:2010}. The purpose of the present paper is to undertake a
systematic exploration of the behaviour of elastic and loss cross sections as a
function of these two parameters and collision energy, and also to compare the
results of the single-channel model with full coupled-channel calculations on a
prototype strongly coupled system, based on rotationally inelastic collisions
of LiH with Li atoms.

\section{Theory}

The single-channel model used here assumes that loss occurs only at short
range, and that flux that leaves the incoming channel does not return. To
calculate the probability of reaching short range, we use a single-channel
Schr\"odinger equation,
\begin{equation}
\left[\frac{-\hbar^2}{2\mu}\frac{d^2}{dr^2}+V(r)+ \frac{\hbar^2 L(L+1)}{2\mu r^2} -E\right]\psi(r ) = 0,
\label{eq:1d}
\end{equation}
where $V(r)$ is the interaction potential, $L$ is the partial-wave quantum
number, and $\mu$ is the reduced mass of the colliding pair. We approximate the
potential $V(r)$ by its long-range form $V_{\rm LR}(r)=-C_6/r^6$; this
simplification allows us to use the analytic QDT formalism of Gao
\cite{Gao:2001, Gao:2008}, which accurately represents the behaviour of the
system across a wide range of energy around threshold \cite{Frye:2014}.
Equation (\ref{eq:1d}) is conveniently rescaled by the van der Waals energy and
length scales $r_{6} = (2\mu C_6/\hbar^2)^{1/4}=2r_{\rm vdw}$ and $E_{6}
=\hbar^2/(2\mu r_{6}^2)$ \cite{Gao:2008} to give
\begin{equation}
\left[-\frac{d^2}{dr_{\rm s}^2}+U(r_{\rm s} )
+ \frac{L(L+1)}{r_{\rm s}^2} -\epsilon\right]\psi(r_{\rm s} ) = 0,
\label{schrod_eqn}
\end{equation}
where $r_{\rm s}=r/r_{6}$, $U(r_{\rm s})=V(r)/E_{6}$ and
$\epsilon=E/E_{6}$. In the present case, $U(r_{\rm s})=-1/r_{\rm s}^6$.

We choose to use the travelling wave reference functions of Sec.\ IIIC and D of
ref.\ \cite{Gao:2008}. These are solutions to equation (\ref{schrod_eqn}) that
have incoming ($-$) or outgoing ($+$) character. They may be normalised in
either the inner region i ($r_{\rm s}\rightarrow 0$) or the outer region o
($r_{\rm s} \rightarrow \infty$),
\begin{eqnarray}
f^{\rm i +}(r_{\rm s}) &\stackrel{r_{\rm s} \rightarrow 0 }{\sim}& r_{\rm s}^{3/2} \exp [-i(r_{\rm s}^{-2}/2-\pi/4)] \\
f^{\rm i -}(r_{\rm s}) &\stackrel{r_{\rm s} \rightarrow 0 }{\sim}& r_{\rm s}^{3/2} \exp [+i(r_{\rm s}^{-2}/2-\pi/4)]\\
f^{\rm o +}(r_{\rm s}) &\stackrel{r_{\rm s} \rightarrow \infty }{\sim}& k_{\rm s}^{-1/2} \exp [+i k_{\rm s} r_{\rm s}]\\
f^{\rm o -}(r_{\rm s}) &\stackrel{r_{\rm s} \rightarrow \infty }{\sim}& k_{\rm s}^{-1/2} \exp [-i k_{\rm s} r_{\rm s}]
\end{eqnarray}
where $k_{\rm s} = \epsilon^{1/2}$. Note that these differ from Gao's
definitions \cite{Gao:2008} by a constant factor of $\pi^{-1/2} e^{i\pi/4}$.
The reference functions in the inner and outer regions are related by
\begin{eqnarray}
f^{\rm o -} + r^{\rm (oi)}f^{\rm o +} = t^{\rm (oi)} f^{\rm i -}
\label{eq:oi}\\
f^{\rm i +} + r^{\rm (io)}f^{\rm i -} = t^{\rm (io)} f^{\rm o +}.
\label{eq:io}
\end{eqnarray}
Equation (\ref{eq:oi}) is interpreted as a wave $f^{\rm o -}$ travelling from
the outer region inwards, which is partially reflected ($r^{\rm (oi)}f^{\rm o
+}$) and partially transmitted ($t^{\rm (oi)} f^{\rm i -}$). Equation
(\ref{eq:io}) is interpreted similarly for the wave $f^{\rm i +}$ travelling in
the opposite direction. The complex coefficients $t$ and $r$ are functions of
$\epsilon$ and $L$, which are readily computed from expressions given in Sec.\
IV of ref.\ \cite{Gao:2008} and Gao's QDT functions $Z^{\rm c}(\epsilon,L)$
\cite{Gao:C6:1998, Gao:AQDTroutines}. The coefficients are related by
\begin{eqnarray}
|r^{\rm (io)}|&=&|r^{\rm (oi)}|\\
t^{\rm (io)}&=&t^{\rm (oi)}\\
|r^{\rm (io)}|^2+|t^{\rm (io)}|^2&=&1.
\end{eqnarray}

The short-range physics is modelled by the boundary condition
\begin{equation}
\psi  \stackrel{r_{\rm s} \rightarrow 0}{\sim} C(f^{\rm i-} + S^{\rm c} f^{\rm i+})
\label{short-range}
\end{equation}
where $C$ is an arbitrary normalisation constant and $S^{\rm c}$ is the
short-range S-matrix. In a true single-channel problem, $S^{\rm c}$ would be a
complex number of magnitude 1, but to account for the loss of flux to other
outgoing channels we allow it to have magnitude $|S^{\rm c}| < 1$. We write
\begin{equation}
S^{\rm c}=\left(\frac{1-y}{1+y}\right)e^{2i\delta^{\rm s}}
\end{equation}
where $y$ is the loss parameter of Idziaszek and Julienne
\cite{Idziaszek:PRL:2010}, so that $y=1$ corresponds to complete loss at short
range and $y=0$ corresponds to no loss. The short-range phase shift
$\delta^{\rm s}$ may be related to the ``background" scattering length $a$
\footnote{The quantity $a$ was termed a background scattering length in ref.\
\cite{Idziaszek:PRL:2010}, but in the presence of closed channels it may
nevertheless contain contributions from Feshbach resonances.} of ref.\
\cite{Idziaszek:PRL:2010},
\begin{equation}
\frac{a}{\bar{a}}=1+\cot\left(\delta^{\rm s}-\frac{\pi}{8}\right),
\end{equation}
where $\bar{a}=0.477 988\ldots r_{6}$ is the mean scattering length of Gribakin
and Flambaum \cite{Gribakin:1993}. This allows us to map the complete range of
behaviours onto the finite range $0 \leq \delta^{\rm s} < \pi$ rather than the
infinite range of $s=a/\bar{a}$ as in ref.\ \cite{Idziaszek:PRL:2010}.

The formulation in terms of $|S^{\rm c}| = (1-y)/(1+y)$ and $\delta^{\rm s}$
makes it obvious that the collisional properties of the system are independent
of $\delta^{\rm s}$ (and hence of $s$) in the limit $y\rightarrow 1$ ($|S^{\rm
c}|\rightarrow 0$). We make the usual QDT approximation that $S^{\rm c}$ is
independent of energy close to threshold, and also that it does not vary with
partial wave $L$ \cite{Gao:2001}.

We obtain the long-range S-matrix $S_L$ for partial wave $L$ by matching
$\psi(r_{\rm s})$, equation (\ref{short-range}), to the usual scattering
boundary conditions,
\begin{equation}
\psi \stackrel{r_{\rm s} \rightarrow \infty }{\sim} f^{\rm o-} - (-1)^L S_L f^{\rm o+}.
\end{equation}
Again, $S_L$ would be unitary in a true single-channel problem but here it can
be sub-unitary. The relationship between $S^{\rm c}$ and $S_L$ is given by
Sec.\ VIIB of ref.\ \cite{Gao:2008} as
\begin{equation}
S_L=(-1)^{L+1} \left[r^{\rm (oi)}+\frac{t^{\rm (oi)} S^{\rm c} t^{\rm (io)}}{1-r^{\rm (io)}S^{\rm c}}\right],
\label{eq:SL}
\end{equation}
which may be expanded as
\begin{eqnarray}
S_L&=&(-1)^{L+1} \bigl[r^{\rm (oi)}\nonumber\\
&+&t^{\rm (oi)} S^{\rm c} t^{\rm (io)} (1  + r^{\rm (io)}S^{\rm c} + (r^{\rm (io)}S^{\rm c})^2 + \dots) \bigr].
\label{eq:SLexp}
\end{eqnarray}
Equation (\ref{eq:SLexp}) provides a clear physical understanding of the
scattering process. It is made up of multiple pathways: reflection off the pure
long-range potential; transmission inwards past the long-range potential,
followed by a single interaction with the short-range and retransmission out
past the long-range potential; then a further series of terms which involve
repeated reflections off the long-range potential back towards short range.
This last group is responsible for shape resonances when $r^{\rm (io)}S^{\rm
c}$ is close to $1$ and successive terms of the sum add constructively. The
resonances are damped if $|S^{\rm c}| < 1$ ($y> 0$). In the limit $S^{\rm c}
\rightarrow 0$ the only elastic scattering is reflection off the long-range
potential, since all flux transmitted past the long-range potential is lost and
not reflected back out.

The total elastic and loss cross sections may both be expressed in terms of the
elastic S-matrix elements,
\begin{eqnarray}
\sigma_{\rm el} &=& \frac{g\pi}{k^2} \sum_L (2L+1) |1-S_L|^2, \\
\sigma_{\rm loss} &=& \frac{g\pi}{k^2} \sum_L (2L+1) (1-|S_L|^2).
\end{eqnarray}
For distinguishable particles, the symmetry factor $g$ is 1 and the sum runs
over all values of $L\ge0$; for identical bosons or fermions, $g$ is 2 and the
sum runs over only even or only odd values of $L$, respectively.

\section{Results of the single-channel model}

\begin{figure*}[tbp]
\includegraphics[height=0.9\textheight]{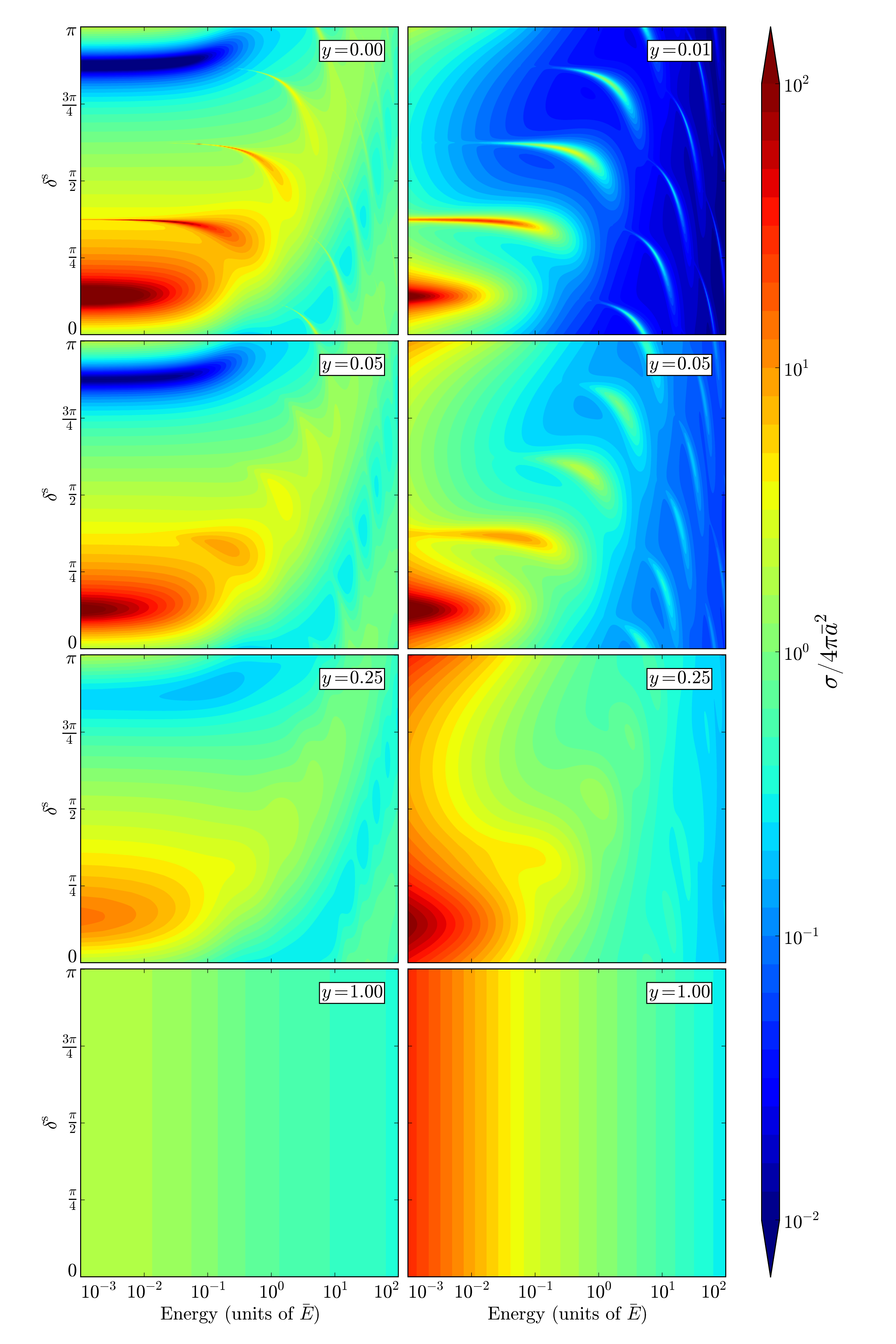}
\caption{Contour plots of the elastic (left) and loss (right) cross sections
for distinguishable particles as a function of reduced energy $E/\bar{E}$ and
short-range phase shift $\delta^{\rm s}$ for selected values of the loss
parameter $y$.} \label{fig:dist_samp}
\end{figure*}

\begin{figure*}[tbp]
\includegraphics[height=0.9\textheight]{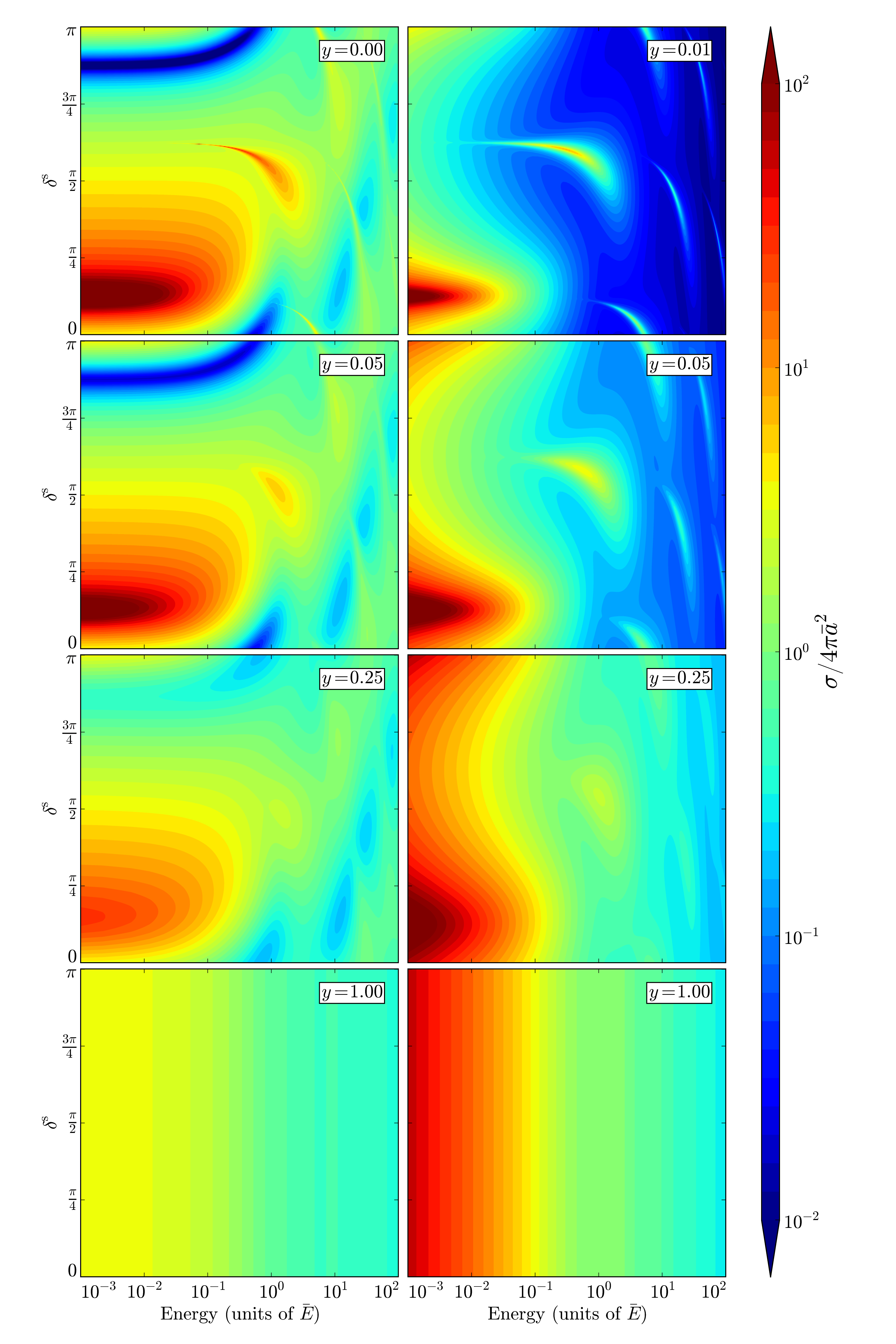}
\caption{Contour plots of the elastic (left) and loss (right) cross sections
for identical bosons as a function of reduced energy $E/\bar{E}$ and
short-range phase shift $\delta^{\rm s}$ for selected values of the loss
parameter $y$.} \label{fig:bos_samp}
\end{figure*}

\begin{figure*}[tbp]
\centering\includegraphics[height=0.9\textheight]{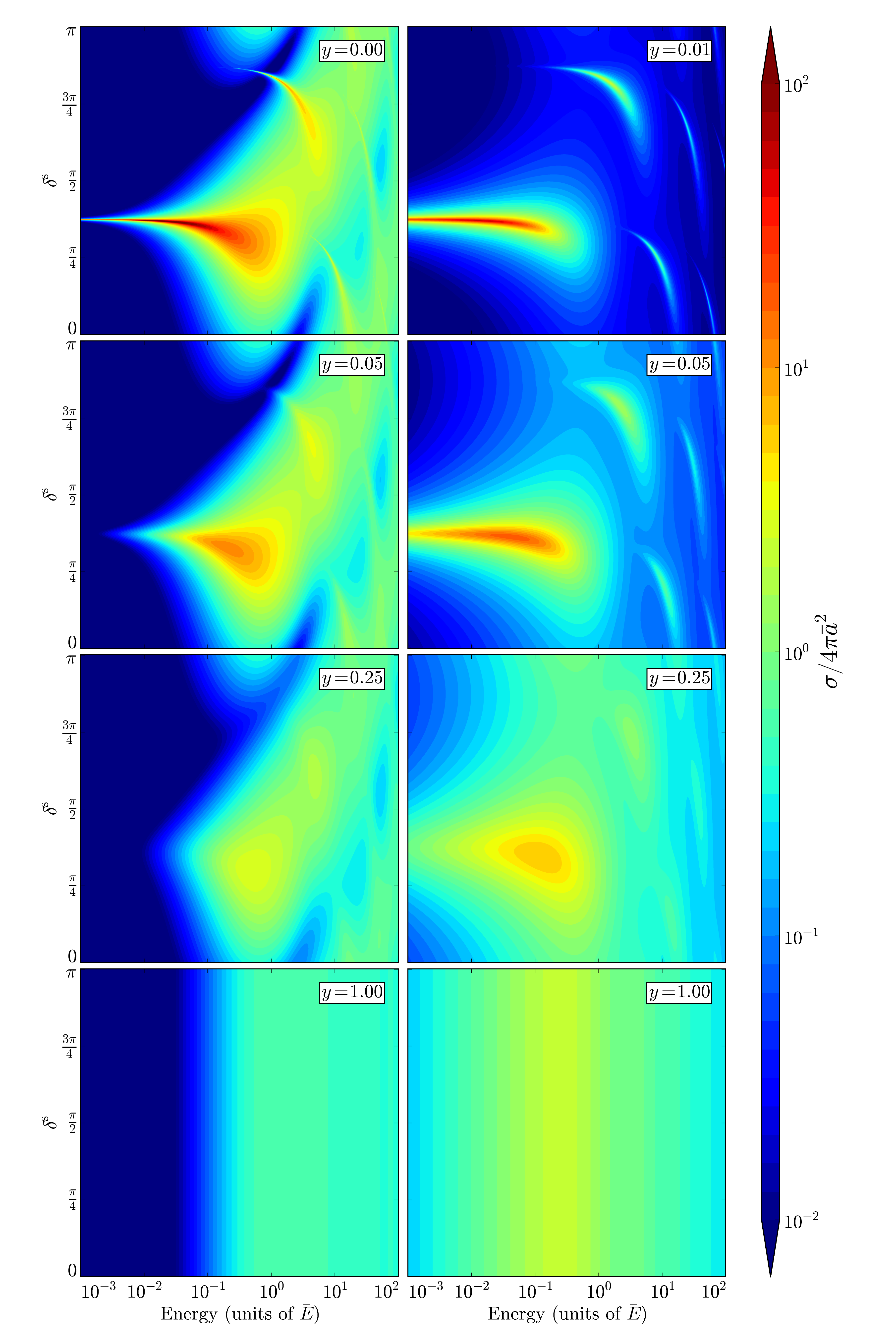}
\caption{Contour plots of the elastic (left) and loss (right) cross sections
for identical fermions as a function of reduced energy $E/\bar{E}$ and
short-range phase shift $\delta^{\rm s}$ for selected values of the loss
parameter $y$.} \label{fig:ferm_samp}
\end{figure*}

Figures \ref{fig:dist_samp} to \ref{fig:ferm_samp} show the elastic and loss
cross sections for selected values of the loss parameter $y$, as a function of
the short-range phase shift $\delta^{\rm s}$ and the reduced energy
$E/\bar{E}$, for distinguishable particles, identical bosons, and identical
fermions, respectively. Figures \ref{fig:dist_all} to \ref{fig:ferm_all} show
animations of the same quantities, with frames at every value of $y$ from 0
(fully elastic) to 1 (the universal loss regime) in steps of 0.01. The
information in these animations is provided in pdf form in Supplementary
Material \footnote{The animations of contour plots in Figs.\ 4, 5 and 6 and
provided in pdf format in Supplementary Material}. The energy scale
$\bar{E}=\hbar^2/(2\mu\bar{a}^2)$ is strongly mass-dependent: for example,
$\bar{E}/k_{\rm B}=61$~mK for He$^*$+Ar \cite{Jachymski:react:2013} but only 97
$\mu$K for KRb+KRb \cite{Kotochigova:react:2010}.

We consider first the case of distinguishable particles. At $y=0$ (Fig.\
\ref{fig:dist_samp}(a)), low-energy scattering is dominated by s-wave features.
There is a large peak near $\delta^{\rm s}=\pi/8$, which corresponds to
infinite scattering length, and a deep trough around $\delta^{\rm s}=7\pi/8$,
which corresponds to zero scattering length. There is a set of sharp shape
resonances that curve towards their zero-energy positions: p-wave at
$\delta^{\rm s}=3\pi/8$, d-wave at $\delta^{\rm s}=5\pi/8$, and further partial
waves at increments of $\pi/4$. Thus a shape resonance in partial wave $L+4$
has the same zero-energy position as that in partial wave $L$ \cite{Gao:2000},
e.g. an h-wave ($L=5$) shape resonance curves towards the same zero-energy
position as the p-wave ($L=1$) resonance. The plots are cyclic in $\delta^{\rm
s}$ with period $\pi$, so that the contours along the top edge of each plot are
the same as those along the bottom edge. It may be noted that the trough
corresponding to zero scattering length curves upwards as a function of energy;
this arises because of a Ramsauer-Townsend minimum \cite{Mott:1965} that occurs
in the s-wave cross section for small negative values of the scattering length.

There is by definition no loss cross section for $y=0$; for $y=0.01$ (Fig.\
\ref{fig:dist_samp}(b)) there is very little loss except close to the shape
resonances: little flux is lost in each interaction with the short-range
region, so it is only at a shape resonance that there are many interactions
with the short-range region and loss becomes important. Shape resonances cause
visible features at least as high as $L=11$ in the plots for $y=0$ and 0.01.
Note that there is a large peak in the s-wave loss near $\delta^{\rm s}=\pi/8$,
even though s-wave collisions cannot have shape resonances {\em per se}: at low
enough energies the long-range potential reflects outgoing flux even with no
barrier, so that the multiple interactions with the short-range region that are
characteristic of shape resonances can still occur.

As the loss parameter $y$ increases from 0, the features in the cross sections
broaden out and eventually disappear, reaching the ``universal loss" regime
described by Idziaszek and Julienne \cite{Idziaszek:PRL:2010}. Most of the
features described above are still visible at $y=0.05$, though the shape
resonances are lower and do not persist to such low energy. However, the
features have largely washed out by $y=0.25$. The amplitude of variations in
$\sigma_{\rm loss}$ as a function of $\delta^{\rm s}$ decreases steadily as $y$
increases. It should be noted that, even though $y=1$ corresponds to complete
loss at short range, it does {\em not} give the maximum possible overall loss
rate. Values of $y<1$ can sometimes give even faster loss rates because of the
possibility of resonant enhancement.

The results for identical bosons (Figs.\ \ref{fig:bos_samp} and
\ref{fig:bos_all}) show similar features to those for distinguishable
particles, except that there are no odd-$L$ shape resonances. However, the
results for identical fermions (Figs.\ \ref{fig:ferm_samp} and
\ref{fig:ferm_all}) are visually very different, because of the lack of an
s-wave background at low energy. The shape resonances (this time for odd $L$
only) are therefore even more prominent.

\begin{figure}[tbp]
\hbox{[Files plots\_d\_el.mov and plots\_d\_ls.mov to be inserted here]}
\caption{Animations of contour plots of the elastic (left) and loss
(right) cross sections for distinguishable particles as a function of reduced
energy $E/\bar{E}$ and short-range phase shift $\delta^{\rm s}$ as the loss
parameter $y$ varies from 0 to 1.} \label{fig:dist_all}
\end{figure}

\begin{figure}[tbp]
\hbox{[Files plots\_b\_el.mov and plots\_b\_ls.mov to be inserted here]}
\caption{Animations of contour plots of the elastic (left) and loss
(right) cross sections for identical bosons as a function of reduced
energy $E/\bar{E}$ and short-range phase shift $\delta^{\rm s}$ as the loss parameter
$y$ varies from 0 to 1.} \label{fig:bos_all}
\end{figure}

\begin{figure}[t]
\hbox{[Files plots\_f\_el.mov and plots\_f\_ls.mov to be inserted here]}
\caption{Animations of contour plots of the elastic (left) and loss
(right) cross sections for identical bosons as a function of reduced
energy $E/\bar{E}$ and short-range phase shift $\delta^{\rm s}$ as the loss parameter
$y$ varies from 0 to 1.} \label{fig:ferm_all}
\end{figure}

\begin{figure*}[tbp]
\includegraphics[height=0.9\textheight]{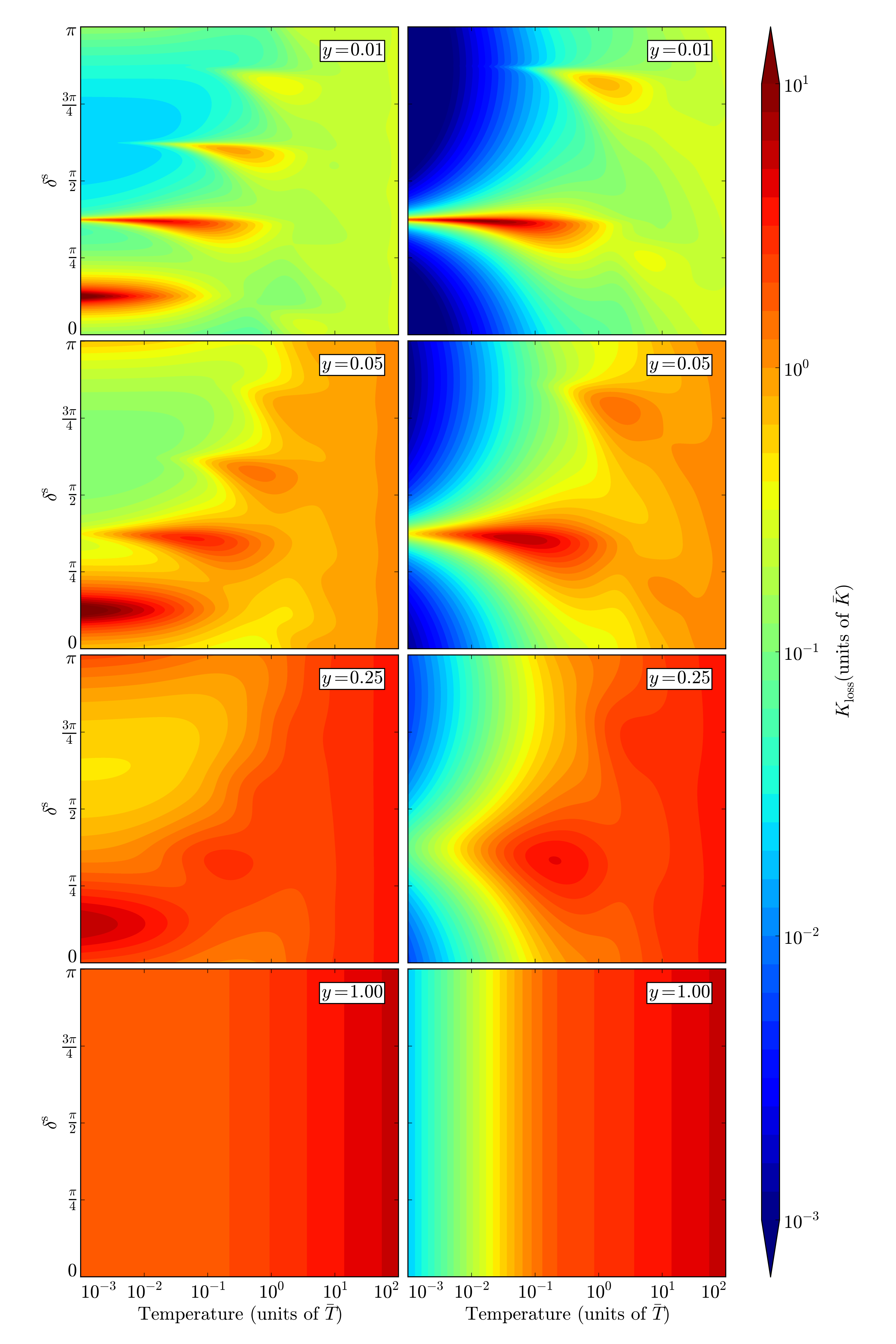}
\caption{Contour plots of the thermally averaged loss rate for distinguishable
particles (left) and identical fermions (right) as a function of reduced
temperature $k_{\rm B}T/\bar{E}$ and short-range phase shift $\delta^{\rm s}$
for selected values of the loss parameter $y$. The loss rate is scaled by
$\bar{K}=\bar{a}h/\mu$.} \label{fig:thermal}
\end{figure*}

Many experiments use thermal samples. Figure \ref{fig:thermal} shows the rate
constants for loss processes as a function of temperature, for selected values
of $y$, for distinguishable particles (left) and identical fermions (right).
The major features of the plots remain, but it may be seen that some of the
higher-energy structure is washed out by averaging over kinetic energy. In
particular, shape resonances due to partial waves with $L>3$ are barely
visible.

\section{Comparison with coupled-channel calculations}

Most of the real collision systems of interest in ultracold physics are
multichannel in nature and have both shape and Feshbach resonances. It is
interesting to consider how far the single-channel model described here can
reproduce the results of full coupled-channel calculations in such systems. To
explore this, we have carried out full coupled-channel calculations on
field-free rotationally inelastic collisions in the system Li+LiH. This is a
strongly coupled system with a highly anisotropic potential energy surface
\cite{Skomorowski:LiHLi:2011}. In previous work, we calculated elastic and
inelastic collision cross sections of $^7$LiH+$^7$Li with the molecules
initially in the ground state and the first rotationally excited state
\cite{Tokunaga:2011}. In the present work, we extend these calculations to
consider LiH molecules initially in higher rotational states $j$, so that there
are many inelastic (loss) channels available. The calculations are carried out
with the MOLSCAT package \cite{molscat:v14}, using the same methods and basis
sets for solving the coupled-channels as described in ref.\
\cite{Tokunaga:2011}. Cross sections are calculated by summing contributions
from partial waves labelled by the total angular momentum $J$; the sum
converges by $J=13$ at collision energies up to 1~K for the initial rotational
states considered here. We use the potential energy surface from ref.\
\cite{Skomorowski:LiHLi:2011}, except that we introduce a scaling factor
$\lambda$ that allows us to sample different possible values of the short-range
phase shift $\delta^{\rm s}$ \footnote{In scaling the potential, we keep the
long-range part fixed (Eq.\ (2) and Table IV of ref.\
\cite{Skomorowski:LiHLi:2011}), and scale the remainder of the potential by a
factor $\lambda$. This ensures that the Van der Waals length and energy do not
vary.}.

\begin{figure*}[t]
\includegraphics[width=0.9\textwidth]{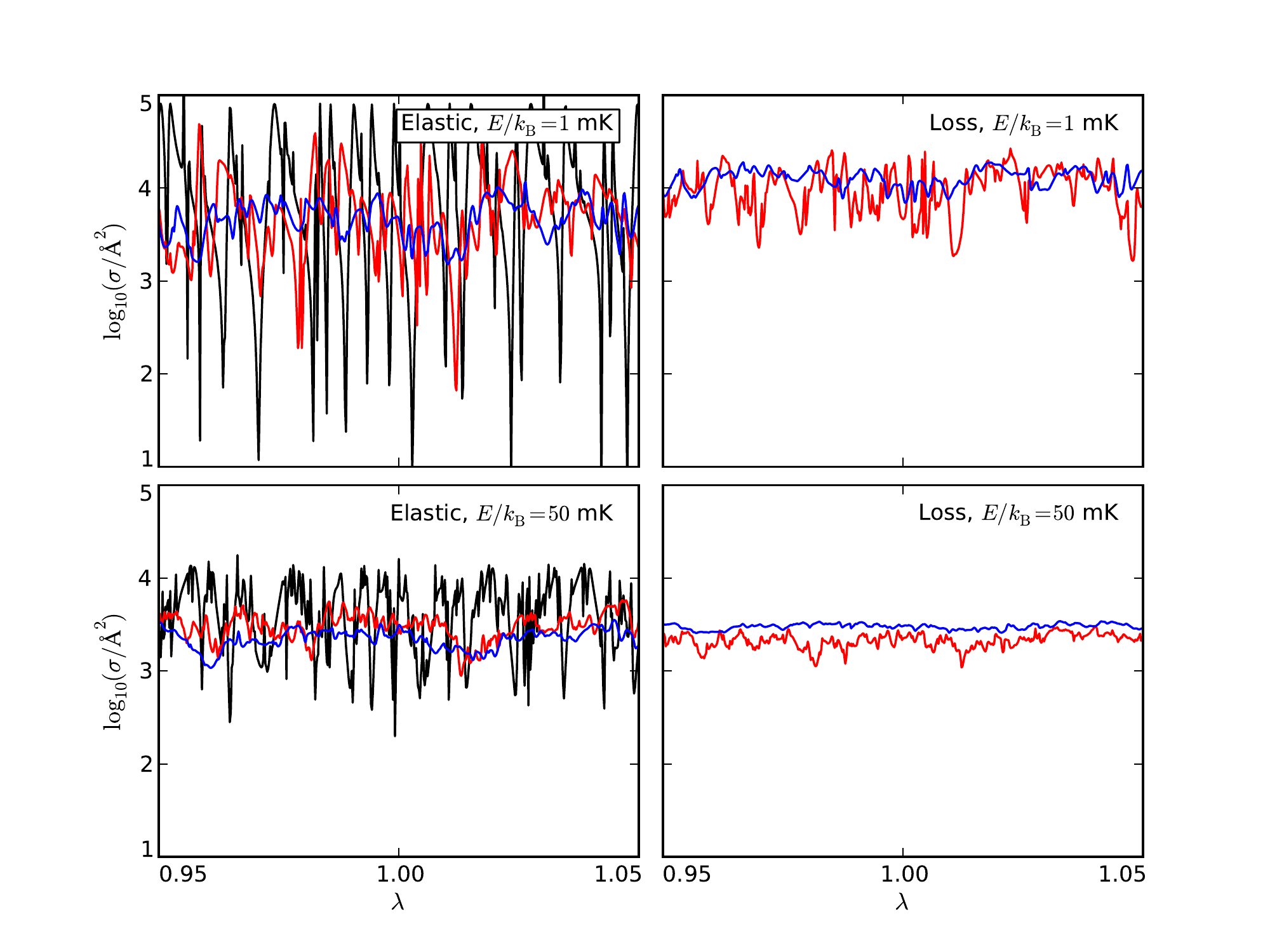}
\caption{Elastic (left) and total inelastic (right) cross sections for Li+LiH
collisions for initial rotational levels $j=0$ (black), 3 (red) and 6 (blue) at
kinetic energies corresponding to $E/k_{\rm B}=1$~mK (top) and $50$ mK (bottom)
as a function of the scaling factor $\lambda$. Note the steadily decreasing
amplitude of oscillations as initial $j$ increases.} \label{fig:vary-lambda}
\end{figure*}

Figure \ref{fig:vary-lambda} shows the calculated elastic and total inelastic
cross sections for Li+LiH collisions for initial rotational levels $j=0$, 3 and
6 at kinetic energies $E/k_{\rm B}=1$~mK and 50~mK as the scaling factor
$\lambda$ is varied across the range $0.95<\lambda<1.05$. The length and energy
scales for Li+LiH are $\bar{a}=16.2$~\AA\ and $\bar{E}/k_{\rm B}=24.5$~mK,
giving a p-wave barrier height of 6.1~mK. Rotationally inelastic collisions are
dominated by couplings at distances much smaller than $\bar{a}$. $E/k_{\rm
B}=1$~mK is in the s-wave regime, so the cross sections for initial $j=0$ show
very large peaks and deep troughs. These correspond to poles and zeroes in the
s-wave scattering length as successive atom-diatom bound states pass through
threshold and cause Feshbach resonances. At $E/k_{\rm B}=50$~mK, peaks and
troughs are still visible, but are less pronounced because of contributions
from higher partial waves and the overall $k^{-2}$ factor in the expressions
for cross sections \cite{Wallis:LiNH:2011}.

For successively higher initial $j$ values, the number of inelastic channels
increases and inelastic scattering becomes progressively stronger. The poles in
scattering length that occur for initial $j=0$ are replaced by finite
oscillations that diminish in amplitude as the inelasticity increases
\cite{Hutson:res:2007}. The amplitude of the oscillations in the cross sections
thus decreases as initial $j$ increases, even in the s-wave regime.

The interaction potential of ref.\ \cite{Skomorowski:LiHLi:2011} has an
estimated uncertainty of only 0.05\%, which is unusually precise for potentials
from electronic structure calculations. In cases where the uncertainty is 1 to
5\%, which is more typical, it is sufficient to span many oscillations in the
cross sections in a plot such as Fig.\ \ref{fig:vary-lambda}. Under these
circumstances it is not meaningful to regard the results of scattering
calculations on a single potential as predictions for the physical system, and
it is essential to understand the {\em range} of results that may be obtained
across the uncertainties in the potential \cite{Hutson:IRPC:2007,
Wallis:MgNH:2009}. It is clear from Fig.\ \ref{fig:vary-lambda} that the range
of possible results is very large for purely elastic collisions in the s-wave
regime, but diminishes both when loss is present (for initial $j>0$) and when
there are significant contributions from several partial waves
\cite{Wallis:LiNH:2011}.

\begin{figure}[tbp]
\includegraphics[width=\columnwidth]{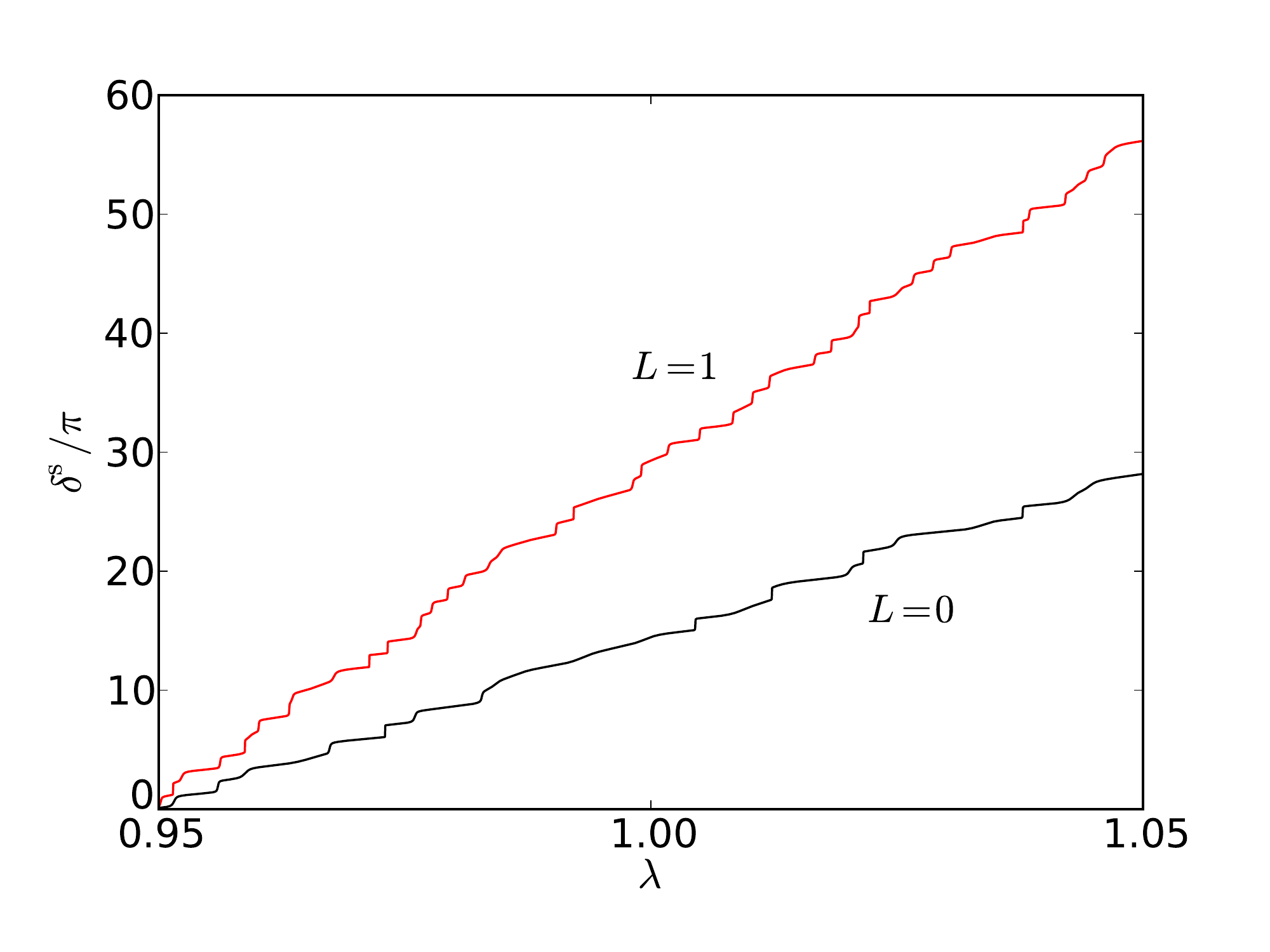}
\caption{Short-range phase shift extracted from the coupled-channel results
at $E/k_{\rm B}=1$ mK, as a function of scaling factor
$\lambda$, for initial $j=0$, $L=0$ (black) and initial $j=0$, $L=1$ (red).}
\label{fig:phase-lambda}
\end{figure}

It is possible to extract values of the short-range phase shift $\delta^{\rm
s}$ and the loss parameter $y$ from coupled-channel results by inverting
equation (\ref{eq:SL}) for a particular channel. The lower (black) line in
Fig.\ \ref{fig:phase-lambda} shows the short-range phase extracted from the
coupled-channel results for initial $j=0$, $L=0$ across the range of scaling
factors $\lambda$ considered. Feshbach resonances appear as an increase of
$\pi$ in $\delta^{\rm s}$ over a small range of $\lambda$. There are 23
resonances of various widths across the range $0.95<\lambda<1.05$, superimposed
on a steadily increasing background. The widths of the resonances are
comparable to their spacings with respect to both energy and $\lambda$ scaling,
so that even s-wave scattering is influenced by resonance effects for most
values of $\lambda$. It may be noted that the corresponding resonances in the
long-range phase shift have widths that are reduced near threshold
\cite{Timmermans:1999} and are strongly energy-dependent. By contrast, the
resonances in the short-range phase shift have widths that are only weakly
energy-dependent and correspond to the widths of the features in low-energy
cross sections.

\begin{figure}[tbp]
\includegraphics[width=\columnwidth]{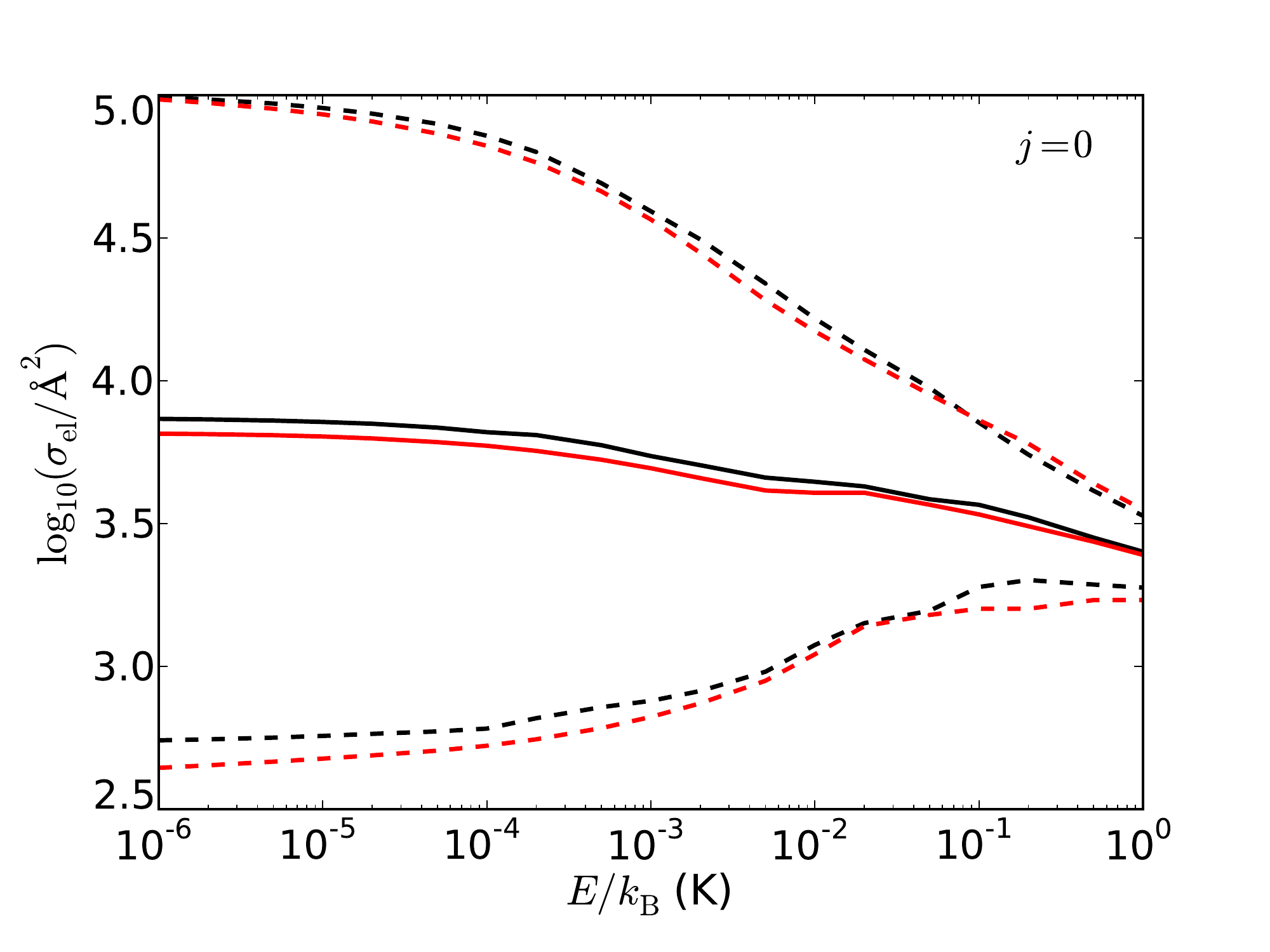}
\caption{Mean values and mean $\pm$ 1 standard deviation of
$\log_{10}(\sigma_{\rm el}$/\AA$^2$) from the single-channel model with $y=0$
(red), compared with the corresponding quantities from coupled-channel
calculations for Li+LiH collisions with initial $j=0$ (black).}
\label{fig:vary-e-j0}
\end{figure}

In a multichannel system, the scattering for $L>0$ is not fully determined by
the value of $\delta^{\rm s}$ obtained for $L=0$. The upper (red) line in Fig.\
\ref{fig:phase-lambda} shows the short-range phase shift obtained by inverting
Eq.\ (\ref{eq:SL}) for initial $j=0$, $L=1$. All the resonances that were
present for $L=0$ appear again, shifted to slightly higher $\lambda$ and often
with somewhat different widths. However, there are 28 additional resonances.
The variation of $\delta^{\rm s}$ with $L$ prevents the single-channel model
giving accurate energy-dependent cross sections for a specific interaction
potential, even for initial $j=0$. In addition, for $j>0$ the value of $y$
obtained by inverting Eq.\ (\ref{eq:SL}) is a fast function of $\lambda$, even
in the s-wave regime, and is also $L$-dependent. Nevertheless, it is useful to
compare the {\em distribution} of elastic and inelastic cross sections obtained
from coupled-channel calculations (as $\lambda$ is varied over the range shown
in the Figures) with that obtained from the single-channel model (as
$\delta^{\rm s}$ is varied from 0 to $\pi$ for a given value of $y$). Figure
\ref{fig:vary-e-j0} shows this comparison for the mean and mean $\pm$ 1
standard deviation of $\log\sigma_{\rm el}$ as a function of collision energy
for the case of initial $j=0$, where there are no inelastic channels, so $y=0$.
It may be seen that the single-channel model (with no adjustable parameters
whatsoever) quite accurately reproduces the energy-dependence of both the mean
and standard deviation, despite the fact that most of the structure in Fig.\
\ref{fig:vary-lambda} comes from Feshbach resonances rather than shape
resonances.

\begin{figure*}[tbp]
\includegraphics[width=0.97\textwidth]{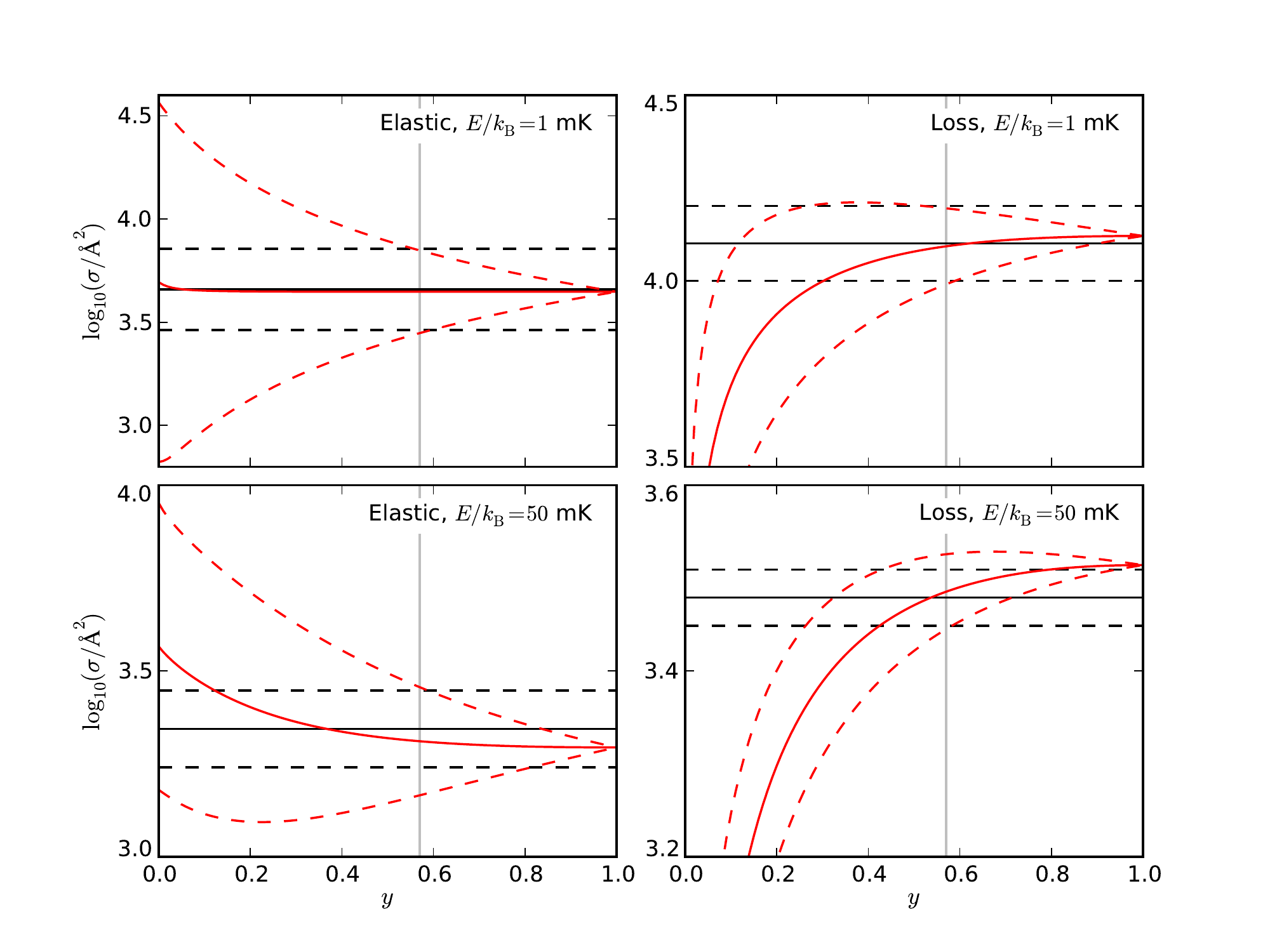}
\caption{Mean values and and mean $\pm$ 1 standard deviation of
$\log_{10}(\sigma_{\rm el}$/\AA$^2$) (left) and $\log_{10}(\sigma_{\rm
loss}$/\AA$^2$) (right) from the single-channel model (red) for collision
energies $E/k_{\rm B}=1$~mK (top) and 50~mK (bottom) as a function of $y$,
compared with the corresponding quantities from coupled-channel calculations
for for Li+LiH collisions with initial $j=6$ (black horizontal lines). The
vertical grey lines indicate $y=0.57$, which gives the best agreement between
the single-channel model and coupled-channel calculations for $j=6$ at low
energy.} \label{fig:vary-y-j6}
\end{figure*}

\begin{figure*}[tbp]
\includegraphics[width=0.97\textwidth]{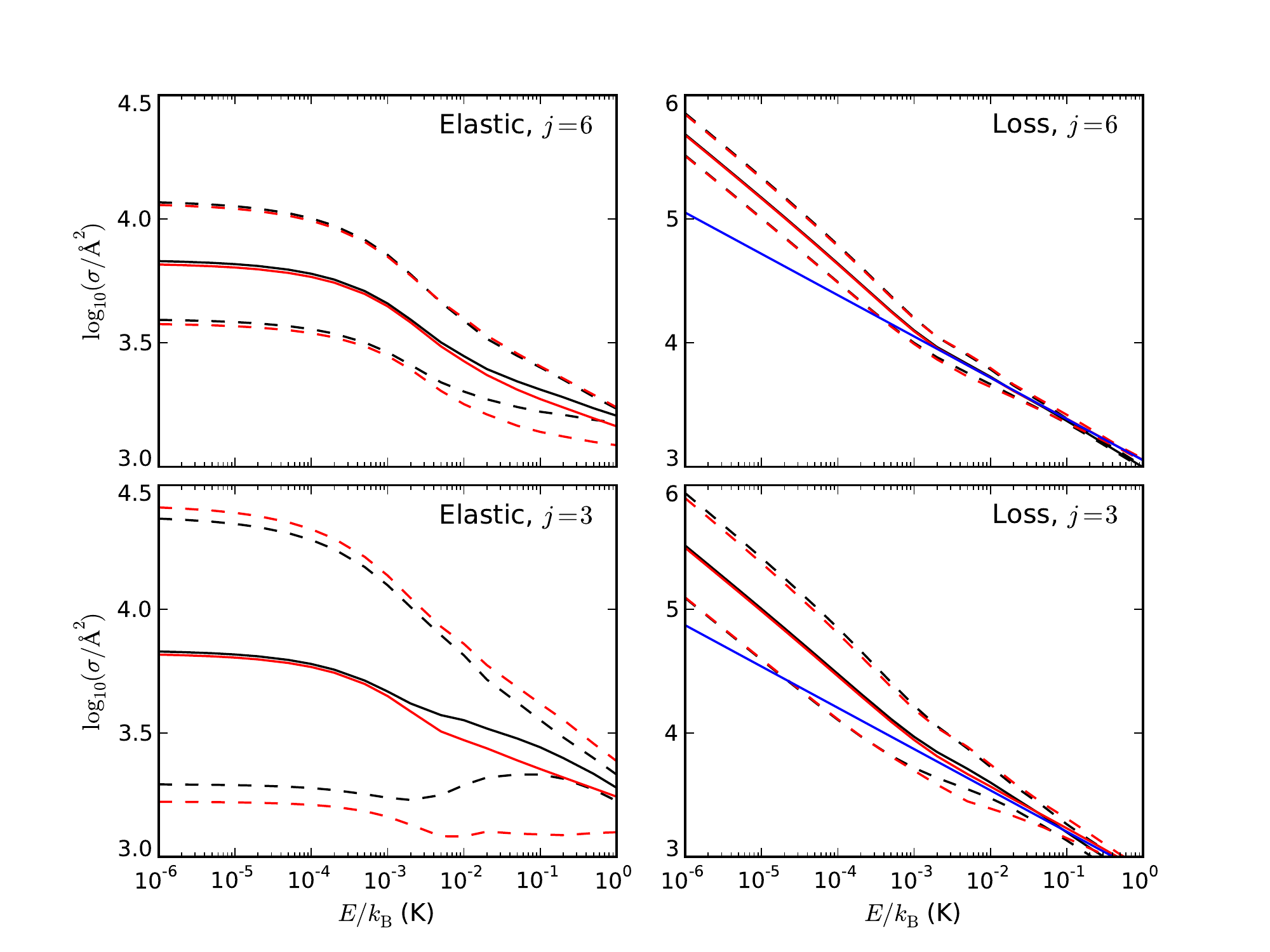}
\caption{Mean values and and mean $\pm$ 1 standard deviation of
$\log_{10}(\sigma_{\rm el}$/\AA$^2$) (left) and $\log_{10}(\sigma_{\rm
loss}$/\AA$^2$) (right) from the single-channel model (red) and coupled-channel
calculations (black) for Li+LiH collisions with initial $j=6$ (top) and $j=3$
(bottom) as a function of collision energy. The single-channel calculations use
$y=0.57$ for $j=6$ and $y=0.23$ for $j=3$. The blue lines show the classical
Langevin cross section multiplied by the reaction probability $P^{\rm
re}=4y/(1+y)^2$.} \label{fig:vary-e-j36}
\end{figure*}

For higher initial $j$, where inelastic scattering is possible, we need to
choose a value of $y$ before comparing the coupled-channel and single-channel
results. The upper panels of Fig.\ \ref{fig:vary-y-j6} show the mean and mean
$\pm$ 1 standard deviation for $\log\sigma_{\rm el}$ and $\log\sigma_{\rm
loss}$ for initial $j=6$ at $E/k_{\rm B}=1$~mK from coupled-channel
calculations (horizontal lines), compared with those calculated from the
single-channel model as a function of $y$ (converging lines). It may be seen
that $y\approx0.57$ approximately reproduces the low-energy distributions. The
lower panels of Fig.\ \ref{fig:vary-y-j6} show the corresponding plots at
50~mK; the single-channel model with $y=0.57$ still reproduces the distribution
of $\sigma_{\rm loss}$ fairly well, and is also qualitatively correct for
$\sigma_{\rm el}$, though it somewhat overestimates the standard deviation in
this case. The full energy-dependence for $y=0.57$ is shown in the upper panels
of Fig.\ \ref{fig:vary-e-j36}; there are quantitative differences, but the
single-channel model is nevertheless remarkably accurate for the distribution
of both elastic and inelastic cross sections over the range of energies shown.
For comparison Fig.\ \ref{fig:vary-e-j36} also shows the classical Langevin
cross section multiplied by the reaction probability $P^{\rm re}=4y/(1+y)^2$
\cite{Jachymski:react:2013}.

The agreement between the coupled-channel calculations and the single-channel
model does deteriorate somewhat for lower values of initial $j$. This is to be
expected, because these cases have fewer open loss channels and it is therefore
more likely that flux that is initially lost from the incoming channel will
subsequently return to it, violating one of the assumptions of the
single-channel model. The lower panels of Fig.\ \ref{fig:vary-e-j36} show the
case of initial $j=3$, where the low-energy distribution is reasonably well
described by $y=0.23$. In this case, however, the higher-energy cross sections
calculated from the single-channel model deviate somewhat from the
coupled-channel results, particularly for the elastic cross sections.
Nevertheless, qualitative agreement remains.

We have verified that the agreement between the coupled-channel calculations
and the single-channel model improves steadily from initial $j=1$ to 6, as the
number of open loss channels increases. Initial $j=1$ is a special case. In the
presence of inelastic scattering, individual Feshbach resonances exhibit both a
peak and a dip in the real and imaginary parts of the complex scattering
length, and hence in the loss cross section \cite{Hutson:res:2007}. When there
is a single dominant loss channel, the dip in the s-wave cross section can be
very deep \cite{Hutson:HeO2:2009} (and reaches $\sigma_{\rm loss}=0$ when there
is only one loss channel). This behaviour skews the distribution of
$\log_{10}\sigma_{\rm loss}$ at the low end, particularly for initial $j=1$.
For higher initial $j$, the effect is reduced by additional loss channels, and
at higher energies it is reduced by contributions from higher partial waves.

\section{Conclusions}

Single-channel models of inelastic and reactive scattering, based on
quantum-defect theory and a single parameter representing short-range loss,
provide a powerful approximate approach to understanding complicated collision
processes at low kinetic energy. We have investigated how these models behave
over the full parameter space of kinetic energy, short-range phase shift (which
maps to background scattering length) and short-range loss parameter $y$. We
have presented animated contour plots that help to understand how the
sensitivity of cross sections to the background scattering length decreases
both as the loss probability increases and with increasing kinetic energy.

We have also carried out coupled-channel calculations on rotationally inelastic
Li+LiH collisions, as a prototype strongly coupled collision system to test the
results of the single-channel model. The low-energy elastic and total inelastic
(loss) cross sections are very sensitive to the short-range potential, and
oscillate very fast as a function of a potential scaling factor $\lambda$.
However, the amplitude of the oscillations decreases as the initial rotational
quantum number $j$ increases, corresponding to an increasing number of loss
channels (and increasing $y$). The amplitude also decreases as the collision
energy increases. The energy dependence of the {\em distribution} of cross
sections $\sigma$, characterised by the mean and standard deviation of
$\log_{10}\sigma$ with respect to variations in the potential, is well
reproduced by the single-channel model for larger values of initial $j$. For
small $j$ the single-channel model is less accurate but still qualitatively
correct.

The present results elucidate the range of behaviour that can be expected for
cold elastic and inelastic (or reactive) collisions in complex systems. They
also demonstrate that single-channel models with a single short-range loss
parameter can correctly reproduce the qualitative features of full-coupled
channel calculations in systems with many open channels, including the
dependence on collision energy. The quality of agreement improves as the
strength of the short-range loss increases. However, specific systems
nevertheless show strong sensitivity to the details of the short-range
interaction potential, which disappears only in the limit of complete
short-range loss.

The present results suggest a remarkable possibility for inferring the
behaviour of cold collisions at higher temperatures from calculations in the
s-wave regime. For a system with enough open channels to be well described by a
single-channel model, it would be possible to perform coupled-channel
calculations for incoming $L=0$ only and use the results (as a function of a
potential scaling factor $\lambda$) to determine a short-range loss parameter
$y$. The single-channel approach could then be used to predict the range of
possible loss rates at higher energy, without the need to carry out explicit
coupled-channel calculations for higher initial $L$.

\begin{acknowledgments}
The authors acknowledge the support of Engineering and Physical Sciences
Research Council Grant no.\ EP/I012044/1, EOARD Grant FA8655-10-1-3033, and
AFOSR-MURI FA9550-09-1-0617.
\end{acknowledgments}

\bibliography{../../all,../universal_mqdt_rates}

\end{document}